\documentclass[aps,prl,reprint,a4paper]{revtex4-1}

\usepackage{amsmath}
\usepackage{amssymb}
\usepackage{graphicx}% Include figure files
\usepackage{dcolumn}% Align table columns on decimal point
\usepackage{bm}% bold math

\usepackage{xcolor}

\begin{document}

%\preprint{}

\title{Energetically-consistent collisional gyrokinetics} %Title of paper



\author{J. W. Burby}
 \affiliation{Princeton Plasma Physics Laboratory, Princeton, New Jersey 08543, USA}
\author{A. J. Brizard}
 \affiliation{Department of Physics, Saint Michael's College, Colchester, Vermont 05439, USA}
\author{H. Qin}
 \affiliation{Princeton Plasma Physics Laboratory, Princeton, New Jersey 08543, USA}
 \affiliation{Dept. of Modern Physics, University of Science and Technology of China, Hefei, Anhui 230026, China}
%\thanks{}


\date{\today}

\begin{abstract}
We present a formulation of collisional gyrokinetic theory with exact conservation laws for energy and canonical toroidal momentum. Collisions are accounted for by a nonlinear gyrokinetic Landau operator. Gyroaveraging and linearization do not destroy the operator's conservation properties. Just as in ordinary kinetic theory, the conservation laws for collisional gyrokinetic theory are selected by the limiting collisionless gyrokinetic theory.
\end{abstract}


\maketitle %\maketitle must follow title, authors, abstract and \pacs

%\emph{Introduction}.\,--- 

\emph{Introduction}.\,--- One of the greatest unsolved problems in the theory of magnetically-confined plasmas is understanding and controlling the turbulent flux of particles and heat into a fusion reactor's wall\,\cite{Kikuchi_2012}. It is believed that the predominant cause of these fluxes is low-frequency fluctuating electromagnetic fields with wavelengths on the order of the gyroradius. While a collisionless gyrokinetic model of these fluctuating fields has been developed that is fully consistent with the First Law of Thermodynamics (for a recent review see Ref.\,\cite{Brizard_2007}), this energetically-consistent model has the serious flaw of ignoring collisions altogether.

In order to accurately describe irreversible plasma transport processes, the effects of collisions must be incorporated into gyrokinetic theory. Previous work on linear gyrokinetic collision operators \cite{Abel_2008,Li_Ernst_2011,Madsen_2013} assumed a strict two-scale separation between a large-scale equilibrium distribution function $F_o$ and a small-scale fluctuating part $\delta F=F-F_o$. Conservation properties of the collision operator in Ref.\,\cite{Abel_2008}, for example, were discussed in the gyroBohm limit. Here, we will focus on nonlinear gyrokinetic collision operators for a global 
full-$F$ approach that do not make this split, and that can thus investigate more completely the possible effects of finite $\epsilon=\rho_i/L$ in experiments, such as corrections to gyroBohm scaling and non-local turbulence spreading (see footnote 5 on p. 427 in Ref.\,\cite{Brizard_2007}.)

%However, because the standard tools of Lagrangian mechanics are inadequate to handle collisional dynamics, the following dichotomy has emerged between collisional and collisionless gyrokinetic theory. Modern collisionless gyrokinetic theory conserves particle number, total energy, and total momentum exactly \cite{Scott_2010}. Modern collisional gyrokinetic theory \cite{Abel_2008,Li_Ernst_2011,Madsen_2013} conserves only particle number. Proposed gyrokinetic collision operators that are purported to conserve energy and momentum \cite{Abel_2008,Li_Ernst_2011} do not in fact lead to gyrokinetic systems of equations with exact energy and momentum conservation laws. 

When finite-$\epsilon$ effects are accounted for, preserving exact conservation properties, and therefore ensuring consistency with the First Law of Thermodynamics, is a nontrivial unsolved problem. The collision operators in Refs.\,\cite{Abel_2008,Li_Ernst_2011}, for example, were obtained by transforming a particle-space collision operator with exact conservation properties into the lowest-order guiding center coordinates. While this approach guarantees the existence of energy and momentum-like quantities that annihilate the collision operator, these same quantities are not conserved by the full-$F$ collisionless gyrokinetic system, and therefore fail to be conserved by the full-$F$ collisional system. More generally, existing gyrokinetic collision operators are not energetically consistent in a full-$F$ formalism because: (a) the gyrocenter coordinate transformation, and therefore any collision operator transformed into gyrocenter coordinates, is only known as an asymptotic expansion in the gyrokinetic ordering parameter $\epsilon$; and (b) replacing the asymptotic expansion of such an operator with a truncated power series destroys exact conservation laws. The purpose of this Letter is to present the first collisional formulation of global full-$F$ gyrokinetics with exact conservation laws. 

\emph{Electrostatic Model}.\,--- For the sake of simplicity, our discussion will focus on quasi-neutral electrostatic gyrokinetics (for instance, see Ref.\,\cite{Parra_2011}). However, the ideas behind our discussion apply equally-well to electromagnetic gyrokinetics (for example, see Ref.\,\cite{Sugama_2000}.) Our primary result consists of an expression for the non-linear Landau operator in gyrocenter coordinates that is corrected by small terms to ensure exact energy and momentum conservation [see Eq.\,(\ref{gyro_landau}).] These correction terms are analogous to the $B_\parallel^*$-denominators in the Hamiltonian guiding center theory introduced by Littlejohn \cite{Littlejohn_1981}; they do not increase the theory's order of accuracy, but they are essential to include for the sake of ensuring exact energy and momentum conservation.

As a first step, we review how the energy conservation law is discussed in collisionless kinetic theory. The governing equations of collisionless electrostatic kinetic theory are the Vlasov-Poisson equations,
\begin{align}
\partial_t f_s+\{f_s,H_s\}=0\label{lp_vlasov_eqn}\\
\Delta\varphi=-4\pi\rho(f), \label{Poisson}
\end{align}
where $f_s$ is the species-$s$ distribution function, $\varphi$ is the electrostatic potential, $\rho(f)$ is the charge density, $H_s= {p^2}/{2m_s}+e_s\varphi$, and $\{\cdot,\cdot\}$ is the standard canonical Poisson bracket. Equations (\ref{lp_vlasov_eqn})-(\ref{Poisson}) conserve the total energy
\begin{align}
{\cal E}&=\sum_{s}\int \frac{p^2}{2 m_s}f_s\,dz+\left\langle\varphi,\frac{}{}\rho(f)+\frac{1}{8\pi}\Delta\varphi\right\rangle,
\label{vp_energy}
\end{align}
where $\langle\cdot,\cdot\rangle$ denotes the standard $L^2$-pairing of functions on configuration space and $dz = d\bm{x}\,d\bm{p}$. Because binary collisions conserve energy, Eq.~(\ref{vp_energy}) must also be conserved in collisional kinetic theory. In particular, if the Vlasov-Poisson equations are modified by the addition of a bilinear collision operator,
\begin{align}
\partial_t f_s+\{f_s,H_s\}=\sum_{\bar{s}} C_{s\bar{s}}(f_s,f_{\bar{s}})\label{landau_poisson_a}\\
\Delta\varphi=-4\pi\rho(f)\label{landau_poisson_b},
\end{align}
then $C_{s\bar{s}}$ must be chosen to satisfy the condition
\begin{align}
0=&\frac{d{\cal E}}{dt} = \sum_{s}\int  H_s\,\partial_t f_s\,dz+\left\langle\partial_t \varphi,\rho(f)+\frac{1}{4\pi}\Delta\varphi\right\rangle\nonumber\\
=&\sum_{s,\bar{s}}\int H_sC_{s\bar{s}}(f_s,f_{\bar{s}})\,dz.
\end{align}
Because this identity must hold for an arbitrary multi-species distribution function, the collision operator therefore has to satisfy the well-known identities
\begin{align}
\int H_sC_{s\bar{s}}(f_s,f_{\bar{s}})\,dz + \int H_{\bar{s}}C_{\bar{s}s}(f_{\bar{s}},f_s)\,d\bar{z} = 0,
\label{basic_identities}
\end{align}
which express the fact that the energy gained by species $s$ due to collisions with species $\bar{s}$ is precisely the energy lost by species $\bar{s}$ due to collisions with species $s$. The non-linear Landau operator (summation rule is implied),
\begin{align}
C_{s\bar{s}}(f_s,f_{\bar{s}})=-\frac{\Gamma_{s\bar{s}}}{2}\{x_i,\gamma_{i}^{s\bar{s}}\},\label{particle_C}
\end{align}
satisfies the identities\,(\ref{basic_identities}), and therefore defines an energetically-consistent collisional kinetic theory. Here $\Gamma_{s\bar{s}}=4\pi e_s^2e_{\bar{s}}^2\ln\Lambda$; the $3$-component vector $\bm{\gamma}^{s\bar{s}}$ is
\begin{align}
\gamma_{i}^{s\bar{s}}(z)=\int \delta(\bm{x}-\bar{\bm{x}})\;\mathbb{Q}^{s\bar{s}}(z,\bar{z})\,\bm{A}_{s\bar{s}}(z,\bar{z})\,d\bar{z};
\end{align}
the $3\times 3$ matrix $\mathbb{Q}^{s\bar{s}}$ is given by
\begin{align}
\mathbb{Q}^{s\bar{s}}(z,\bar{z})=\frac{1}{W_{s\bar{s}}(z,\bar{z})}\mathbb{P}[\bm{W}_{s\bar{s}}(z,\bar{z})],
\end{align}
where $\mathbb{P}(\bm{\xi}) \equiv \mathbb{I} - \hat{\bm{\xi}}\hat{\bm{\xi}}$ is the orthogonal projection onto the plane perpendicular to the vector 
$\bm{\xi}$; the velocity difference $\bm{W}_{s\bar{s}}$ is given by
\begin{align}
\bm{W}_{s\bar{s}}(z,\bar{z})=\{\bm{x},H_s\}(z)-\{\bm{x},H_{\bar{s}}\}(\bar{z});
\end{align}
and the vector 
\begin{align}
\bm{A}_{s\bar{s}}(z,\bar{z})=f_{s}(z)\{\bm{x},f_{\bar{s}}\}(\bar{z})-\{\bm{x},f_s\}(z)f_{\bar{s}}(\bar{z}).
\end{align}
When comparing this form of the Landau operator to more conventional expressions, it is useful to note that $\{\bm{x},g\}=\partial_{\bm{p}}g$, where $g$ is any function on phase space, so that the collision operator (\ref{particle_C}) describes collisions in momentum space. Moreover, the identities
(\ref{basic_identities}) follow immediately from the fact that the velocity difference $\bm{W}_{s\bar{s}}$ is a null-eigenvector of the matrix 
$\mathbb{Q}^{s\bar{s}}$.

\emph{Electrostatic Gyrokinetic Model}.\,--- In order to apply this same argument to gyrokinetic theory, we start with the gyrokinetic Vlasov-Poisson system
\begin{align}
\partial_t F_s+\{F_s,H_s^{\text{gy}}\}_s^{\text{gc}}=0\label{gk_vlasov}\\
\nabla\cdot\bm{P}=\rho(F).\label{qn}
\end{align}
Here, $F_s$ is the gyrocenter distribution function; $\varphi$ is the electrostatic potential; $\{\cdot,\cdot\}_s^{\text{gc}}$ is the guiding center Poisson bracket;
\begin{align}
H_s^{\text{gy}}&=H_s^{\text{gc}}+e_s\left<\psi\right>+\frac{e_s^2}{2}\langle\{\tilde{\psi},\tilde{\Psi}\}_s^{\text{gc}}\rangle \equiv K_s(\bm{E})+e_s\varphi
\end{align}
is the gyrocenter Hamiltonian; $\psi(z)=\varphi(\bm{X}+\bm{\rho}_{os})$, where $\bm{\rho}_{os}$ is the lowest-order guiding-center gyroradius; 
$\langle\cdot\rangle$ denotes the gyroaverage; $\tilde{\Psi}$ denotes the gyroangle antiderivative of $\tilde{\psi} \equiv \psi - \langle\psi\rangle$; $K_s(\bm{E})$ is the gyrocenter kinetic energy; $\bm{P}=-\,
\delta\mathcal{K}/\delta\bm{E}$ is the gyrocenter polarization density; $\mathcal{K}=\sum_s\int F_sK_s(\bm{E})\,dz_s^{\text{gc}}$; and $dz_s^{\text{gc}}$ denotes the guiding center Liouville volume element. These equations govern collisionless quasineutral electrostatic gyrokinetic theory in the ``high-flow" regime (see \cite{Krommes_Hammett_2013} and references therein) and they conserve the total energy,
\begin{align}
{\cal E}^{\text{gy}}=\sum_s\int F_s H_s^{\text{gy}}\,dz_s^{\text{gc}},
\end{align}
exactly. Note that the quasineutrality equation\,(\ref{qn}) implies that this system governs plasma dynamics on time scales long compared to the period of plasma oscillations.

The equations governing collisional gyrokinetic theory are given by adding a bilinear collision operator to the gyrokinetic Vlasov-Poisson equations, 
\begin{align}\label{gk_landau_poisson}
\partial_t F_s+\{F_s,H_s^{\text{gy}}\}_s^{\text{gc}}&=\sum_{\bar{s}}C^{\text{gy}}_{s\bar{s}}(F_s,F_{\bar{s}})\\
\nabla\cdot\bm{P}&=\rho(F).
\end{align}
Because the conservation laws of ordinary collisional kinetic theory are consistent with those of collisionless kinetic theory, the gyrokinetic collision operator $C^{\text{gy}}_{s\bar{s}}$ must not alter the conservation of ${\cal E}^{\text{gy}}$. Thus, 
\begin{align}
0=&\frac{d{\cal E}^{\text{gy}}}{dt}=\sum_s\int H_s^{\text{gy}}\partial_tF_s\,dz_s^{\text{gc}}+\left\langle\rho(F)-\nabla\cdot\bm{P},\frac{}{}\partial_t\varphi\right\rangle\nonumber\\
=&\sum_{s,\bar{s}}\int H_s^{\text{gy}}C^{\text{gy}}_{s\bar{s}}(F_s,F_{\bar{s}})\,dz_s^{\text{gc}}.
\end{align}
This identity will be satisfied for a general multi-species gyrocenter distribution function if and only if
\begin{align}
\int H_s^{\text{gy}}C_{s\bar{s}}^{\text{gy}}(F_s,F_{\bar{s}})\,dz_s^{\text{gc}}+\int H_{\bar{s}}^{\text{gy}}C_{\bar{s}s}^{\text{gy}}(F_{\bar{s}},F_s)\,d\bar{z}_{\bar{s}}^{\text{gc}}=0,\label{gk_identity}
\end{align}
which is the gyrokinetic version of Eq.\,(\ref{basic_identities}). The identities\,(\ref{gk_identity}) must be satisfied exactly by any energetically-consistent gyrokinetic collision operator.

\emph{An energetically-consistent collision operator\,---\,} While Eq.\,(\ref{gk_identity}) imposes important qualitative constraints, they cannot determine the form of the gyrokinetic collision operator by themselves.  A quantitative constraint is necessary as well. To this end, it is important that the gyrokinetic collision operator agrees with the the transformation of the particle-space Landau operator \footnote{Necessary conditions for the use of the Landau operator are $\omega_c<\omega_p$ and $(\partial_tF)/(\omega_p F)<1$. When these conditions are not satisfied, our discussion must be modified.} into gyrocenter coordinates, at least up to some desired order in the gyrokinetic ordering parameter $\epsilon$. Is it possible to satisfy these qualitative and quantitative constraints simultaneously? The answer is ``yes". 

We have discovered an accurate gyrokinetic collision operator that is consistent with the conservation laws of collisionless gyrokinetic theory, and therefore the first law of thermodynamics. The form of the operator is suggested by the somewhat-peculiar presentation of the particle-space Landau operator given earlier. Let $\bm{y}_s=\bm{X}+\bm{\rho}_{os}$ and define the gyrocenter velocity difference
\begin{align}\label{gy_vel_dif}
\bm{W}_{s\bar{s}}^{\text{gy}}(z,\bar{z})=\{\bm{y}_s,H_s^{\text{gy}}\}^{\text{gc}}_s(z)-\{\bm{y}_{\bar{s}},H_{\bar{s}}^{\text{gy}}\}^{\text{gc}}_{\bar{s}}(\bar{z}),
\end{align}
the associated $3\times 3$ matrix
\begin{align}
\mathbb{Q}^{s\bar{s}}_{\text{gy}}(z,\bar{z})=\frac{1}{W^{\text{gy}}_{s\bar{s}}(z,\bar{z})}\mathbb{P}[\bm{W}^{\text{gy}}_{s\bar{s}}(z,\bar{z})],
\end{align}
and the vector
\begin{align}
\bm{A}_{s\bar{s}}^{\text{gy}}(z,\bar{z})=F_{s}(z)\{\bm{y}_{\bar{s}},F_{\bar{s}}\}^{\text{gc}}_{\bar{s}}(\bar{z})-\{\bm{y}_s,F_s\}^{\text{gc}}_s(z)F_{\bar{s}}(\bar{z}).
\end{align}
The energetically-consistent gyrokinetic Landau operator is given by
\begin{align}\label{gyro_landau}
C_{s\bar{s}}^{\text{gy}}(F_s,F_{\bar{s}})=-\frac{\Gamma_{s\bar{s}}}{2}\{y_{s\,i},\gamma^{s\bar{s}}_{\text{gy}\,i}\}^{\text{gc}}_s,
\end{align}
where
\begin{align}
\bm{\gamma}^{s\bar{s}}_{\text{gy}}(z)=\int \delta^{\text{gy}}_{s\bar{s}}(z,\bar{z})\mathbb{Q}^{s\bar{s}}_{\text{gy}}(z,\bar{z})
\bm{A}_{s\bar{s}}^{\text{gy}}(z,\bar{z})\,d\bar{z}_{\bar{s}}^{\text{gc}},
\end{align}
and $\delta^{\text{gy}}_{s\bar{s}}(z,\bar{z})=\delta(\bm{y}_s(z)-\bm{y}_{\bar{s}}(\bar{z}))$. Note that this operator depends explicitly on the electric field through the gyrocenter Hamiltonians that appear in Eq.~(\ref{gy_vel_dif}). Using a straightforward, but tedious argument that is not reproduced here, we have shown that this operator agrees with the Landau operator transformed into gyrocenter coordinates with leading-order accuracy. 

Because the proof is simple, we will now show explicitly that the gyrokinetic Landau-Poisson system\,(\ref{gk_landau_poisson}) defined in terms of the collision operator\,(\ref{gyro_landau}) has exact conservation laws for energy and momentum. We hope to convey the similarity of this demonstration with the analogous demonstration for the ordinary Landau-Poisson system\,(\ref{landau_poisson_a})-(\ref{landau_poisson_b}). However, a word of caution is in order here. It is essential that the guiding center Poisson brackets that appear in Eq.~(\ref{gyro_landau}) be genuine Poisson brackets (i.e., the brackets must satisfy the Leibniz and Jacobi identities). Dropping terms from a bracket that satisfies these properties will destroy the gyrokinetic Landau-Poisson system's exact conservation laws.

\emph{Energy conservation\,---\,}Proving that the gyrokinetic Landau operator\,(\ref{gyro_landau}) satisfies the identities\,(\ref{gk_identity}) is very similar to proving that the particle-space Landau operator satisfies the identities\,(\ref{basic_identities}). Setting $\dot{\mathcal{E}}_{s\bar{s}}=\int H_s^{\text{gy}}C_{s\bar{s}}^{\text{gy}}(F_s,F_{\bar{s}})\,dz^{\text{gc}}_s$, it is simple to verify that
\begin{align}
&\dot{\mathcal{E}}_{s\bar{s}}+\dot{\mathcal{E}}_{\bar{s}s}=\frac{\Gamma_{s\bar{s}}}{2}\iint(\bm{W}_{s\bar{s}}^{\text{gy}})^{\dagger}\mathbb{Q}^{s\bar{s}}_{\text{gy}}\bm{A}_{s\bar{s}}^{\text{gy}}\delta_{s\bar{s}}^{\text{gy}}\,d\bar{z}_{\bar{s}}^{\text{gc}}\,dz_s^{\text{gc}},
\end{align}
where all two-point quantities in the integrand are evaluated at $(z,\bar{z})$ and $\cdot^\dagger$ denotes the ordinary matrix transpose. Because $\mathbb{Q}^{s\bar{s}}_{\text{gy}}$ is a symmetric matrix with null eigenvector $\bm{W}_{s\bar{s}}^{\text{gy}}$, the right-hand-side of this equation vanishes exactly. Thus the gyrokinetic Landau operator\,(\ref{gyro_landau}) satisfies the identities\,(\ref{gk_identity}) exactly, and the gyrokinetic Landau-Poisson system\,(\ref{gk_landau_poisson}) has an exact energy conservation law, $d{\cal E}^{\text{gy}}/dt=0$.

\emph{Toroidal momentum conservation\,---\,}We will prove that if the background magnetic field is axisymmetric, then the gyrokinetic Landau-Poisson system conserves the total toroidal momentum
\begin{align}
P_\phi=\sum_s\int p_{\phi s}F_s\,dz_s^{\text{gc}},\label{gy_Pphi}
\end{align}
where $p_{\phi s}$ is the guiding center canonical toroidal momentum \footnote{Rather than give an explicit expression for $p_{\phi s}$, which will depend on ones choice of guiding center representation, it is better to define it operationally \emph{via} the guiding center Poisson bracket: for each phase space function $f$, the canonical toroidal momentum satisfies $\{f,p_{\varphi s}\}_s^{\text{gc}}=\partial_\phi f$, where $\partial_\phi$ is the toroidal angle derivative.}. If the background magnetic field has additional symmetries, a similar proof of the conservation of the corresponding total momentum can easily be constructed. The time derivative of Eq.~(\ref{gy_Pphi}) yields
\begin{align}
\frac{dP_\phi}{dt}&=\sum_{s,\bar{s}}\int p_{\phi s}C_{s\bar{s}}^{\text{gy}}(F_s,F_{\bar{s}})\,dz_s^{\text{gc}}=\sum_{s,\bar{s}}\dot{P}_{\phi s\bar{s}},
\end{align}
where $P_\phi$ is conserved exactly by the gyrokinetic Vlasov-Poisson system. Here, we find
\begin{align}
&\dot{P}_{\phi s\bar{s}}+\dot{P}_{\phi\bar{s}s}=\nonumber\\
&\frac{\Gamma_{s\bar{s}}}{2}\iint(\{\bm{y}_s,p_{\phi s}\}^{\text{gc}}_s-\{\bm{y}_{\bar{s}},p_{\phi\bar{s}}\}^{\text{gc}}_{\bar{s}})^\dagger\mathbb{Q}^{s\bar{s}}_{\text{gy}}\bm{A}_{s\bar{s}}^{\text{gy}}\delta_{s\bar{s}}^{\text{gy}}\,d\bar{z}_{\bar{s}}^{\text{gc}}\,dz_s^{\text{gc}}.
\end{align}
Now using the fact that $p_{\phi s}$ is the generator of infinitesimal toroidal rotations, we can see that $\{\bm{y}_s,p_{\phi s}\}^{\text{gc}}_s=e_z\times\bm{y}_s$, where $e_z$ is the unit vector along the axis of rotation. Therefore the vector quantity $(\{\bm{y}_s,p_{\phi s}\}^{\text{gc}}_s-\{\bm{y}_{\bar{s}},p_{\phi\bar{s}}\}^{\text{gc}}_{\bar{s}})\,\delta_{s\bar{s}}^{\text{gy}}=e_z\times(\bm{y}_s-\bm{y}_{\bar{s}})\,\delta_{s\bar{s}}^{\text{gy}}=0$, which follows from standard $\delta$-function properties. This shows that $\dot{P}_{\phi s\bar{s}}+\dot{P}_{\phi\bar{s}s}=0$, which in turn implies total toroidal momentum conservation $dP_{\phi}/dt=0$.

\emph{Entropy production\,---\,}As we have discussed, these conservation laws ensure that the gyrokinetic Landau-Poisson system is consistent with the the First Law of Thermodynamics. On the other hand, they do not directly imply that the gyrokinetic Landau-Poisson system is consistent with the Second Law of Thermodynamics. To verify that entropy is indeed a non-decreasing function of time, we have computed the time 
derivative of $S=-\sum_s\int F_s\text{ln} F_s\,dz_s^{\text{gc}}$ and found  
\begin{align}
\frac{dS}{dt}=\frac{\Gamma_{s\bar{s}}}{2}\iint \frac{1}{F_sF_{\bar{s}}}(\bm{A}_{s\bar{s}}^{\text{gy}})^\dagger\mathbb{Q}^{s\bar{s}}_{\text{gy}}\bm{A}_{s\bar{s}}^{\text{gy}}\delta_{s\bar{s}}^{\text{gy}}\,d\bar{z}_{\bar{s}}^{\text{gc}}\,dz_s^{\text{gc}}.\label{S_dot}
\end{align} 
Because $\mathbb{Q}^{s\bar{s}}_{\text{gy}}$ is a positive semi-definite matrix and the distribution function is positive \footnote{Positivity of the distribution function is also guaranteed by the positive semi-definiteness of $\mathbb{Q}^{s\bar{s}}_{\text{gy}}$.}, the right-side of Eq.~(\ref{S_dot}) is non-negative, which is the desired result.

Note that this proves one ``half" of a gyrokinetic version of Boltzmann's $H$-theorem. The missing ingredient is a complete characterization of the distributions that satisfy $dS/dt=0$, i.e. the gyrokinetic Maxwellians. Because the guiding center Poisson bracket is rather complicated, we have not yet found a complete characterization. However, we have verified that the distribution
\begin{align}
F_{Ms}=\frac{1}{Z_s}\exp\bigg(-\frac{H_{s}^{\text{gy}}}{T}\bigg),
\end{align}
where $Z_s=\int \exp(-H_s^{\text{gy}}/T)\,dz_s^{\text{gc}}$ is the partition function, maximizes the entropy. We leave the characterization of the most general gyrokinetic Maxwellian, which would be useful for the sake of deriving dissipative gyrofluid models with exact conservation laws \cite{Madsen_gyrofluid_2013}, as a topic for future study.

\emph{Gyroaveraging\,---\,} When the collision frequency is much smaller than the gyrofrequency \cite{Brizard_2004}, the full gyrokinetic Landau operator\,(\ref{gyro_landau}) can be replaced with that operator's gyroaverage, $\langle C_{s\bar{s}}^{\text{gy}}\rangle$. When this is done, the gyrokinetic Landau-Poisson system becomes the gyroaveraged Landau-Poisson system,
\begin{align}
\partial_t F_s+\{F_s,H_s^{\text{gy}}\}_s^{\text{gc}}&=\sum_s\langle C^{\text{gy}}_{s\bar{s}}(F_s,F_{\bar{s}})\rangle\\
\nabla\cdot\bm{P}&=\rho(F),
\end{align} 
where $F_s$ is now interpreted as the gyroaveraged part of the distribution function. Because the functions $H_{s}^{\text{gy}}$ and $p_{\phi s}$ are independent of the gyrophase, the proofs of energy and momentum conservation given earlier work with $C_{s\bar{s}}^{\text{gy}}$ replaced by $\langle C_{s\bar{s}}^{\text{gy}}\rangle$. Thus, the gyroaveraged Landau-Poisson system has exact energy and momentum conservation laws.

\emph{Linearization\,---\,} Closely related to the gyroaveraged Landau-Poisson system is the collisionally-linear gyroaveraged Landau-Poisson system, 
\begin{align}
\partial_t  F_s+\{ F_s,H_s^{\text{gy}}\}_s^{\text{gc}}&=\sum_{\bar{s}}\left(\delta C_{s\bar{s}}^{\text{test}} \;+\frac{}{}
\delta C_{s\bar{s}}^{\text{field}}\right), \\
\nabla\cdot\bm{P}&=\rho(F),
\end{align} 
where the linearized test-particle and field-particle collision operators are
\begin{align}
\delta C_{s\bar{s}}^{\text{test}}(F_s)&=\langle C^{\text{gy}}_{s\bar{s}}(F_s,F_{M\bar{s}})\rangle, \label{deltaC_test} \\
\delta C_{s\bar{s}}^{\text{field}}( F_{\bar{s}})&=\langle C^{\text{gy}}_{s\bar{s}}( F_{Ms}, F_{\bar{s}})\rangle. \label{deltaC_field}
\end{align}
This system of equations is obtained from the gyroaveraged Landau-Poisson system by assuming $F_s=F_{Ms}+\delta F_{s}$ and then dropping the non-linear term in the collision operator, $\langle C_{s\bar{s}}^{\text{gy}}(\delta F_s,\delta F_{\bar{s}})\rangle$. Note that $\langle C_{s\bar{s}}^{\text{gy}}( F_{Ms}, F_{M\bar{s}})\rangle=0$ \footnote{Note that this identity does not contradict the message presented in Ref.\,\cite{Madsen_2013}. In that reference, the gyrokinetic Maxwellian is defined using only the lowest-order gyrocenter Hamiltonian.}.  Because the gyrokinetic Landau operator satisfies the identities\,(\ref{gk_identity}), it is straightforward to prove that these equations have the same conservation laws for energy and momentum as the gyroaveraged Landau-Poisson system.

\emph{Concluding remarks\,---\,} The key to deriving an energetically-consistent formulation of collisional gyrokinetics was first expressing the particle-space Landau operator in terms of Poisson brackets ``as much as possible," which was an idea first championed by Brizard in Ref.\,\cite{Brizard_2004}. In particular, the identity
\begin{align}
\bm{v}-\bar{\bm{v}}=\{\bm{x},H_s\}(z)-\{\bm{x},H_{\bar{s}}\}(\bar{z})
\end{align}
suggests that the appropriate definition of the gyrocenter velocity difference is given by Eq.\,(\ref{gy_vel_dif}). This idea, together with the procedure given earlier for determining the energetic consistency constraints, appears to be appropriate for deriving energetically-consistent collision operators for other reduced plasma models as well. In future work, we will report on the energy-conserving collisional formulations of electromagnetic gyrokinetics and oscillation center theory. 

We note that, although the gyrokinetic Landau operator (\ref{gyro_landau}) and its linearized forms (\ref{deltaC_test})-(\ref{deltaC_field}) may prove difficult to implement numerically, they identify the proper formalism for the inclusion of collisional transport in gyrokinetic theory. Hence, these gyrokinetic collision operators form the basis from which approximations can be implemented for practical applications.

Lastly, by setting $\varphi=0$ in the above formulas, our results reduce to an energy-momentum-conserving guiding center collision operator. This operator would be ideally suited to incorporating collisions into orbit-following codes such as ORBIT\,\cite{White_1984}; see Ref.\,\cite{Hirvijoki_2013} for recent work on the Monte Carlo implementation of a 5D guiding center Fokker-Planck collision operator. All previous guiding center collision operators that have been applied in orbit-following codes either resort to \emph{ad hoc} methods to ensure exact conservation laws \cite{Boozer_collisions_1981}, or else inconsistently account for inhomogeneities in the magnetic field\,\cite{Tessarotto_1994}.
\begin{acknowledgments}
% Put your acknowledgments here.
This work was supported by DOE contracts DE-AC02-09CH11466 (JWB and HQ) and DE-SC0006721 (AJB).
\end{acknowledgments}

%\appendix
%\section{\label{appA}}


%\bibliography{cumulative_bib_file}

%%%%%%%%%%%%%%%%%%%%%%%%%%%%%%%%%%%%%

%% put content of bib file here when ready to submit

\providecommand{\noopsort}[1]{}\providecommand{\singleletter}[1]{#1}%
%

%%%%%%%%%%%%%%%%%%%%%%%%%%%%%%%%%%%%

\end{document}